\documentclass[twocolumn, secnumarabic, amssymb, nobibnotes, aps, prab, amsmath]{revtex4-1}
\usepackage[english]{babel}
\usepackage{graphicx}

\begin{document}

\title{%
Diffraction at the Open-Ended Dielectric-Loaded \\ Circular Waveguide:
Rigorous Approach%
 }
\author{Sergey N. Galyamin}
\email{s.galyamin@spbu.ru}
\author{Viktor V. Vorobev}
\author{Andrey V. Tyukhtin}

\affiliation{Saint Petersburg State University, 7/9 Universitetskaya nab., St. Petersburg, 199034 Russia}


\begin{abstract}
An elegant and convenient rigorous approach for solving circular open-ended dielectric-loaded waveguide diffraction problems is presented.
It uses the solution of corresponding Wiener-Hopf-Fock equation and leads to an infinite linear system for reflection coefficients (S-parameters) of the waveguide, the latter can be efficiently solved numerically using the reducing technique.
As a specific example directly applicable to beam-driven radiation sources based on dielectric-lined capillaries, diffraction of a slow TM symmetrical mode at the open end of a circular waveguide with uniform dielectric filling is considered.
A series of such modes forms the wakefield (Cherenkov radiation field) generated by a charged particle bunch during its passage along the waveguide axis.
Calculated S-parameters were compared with those obtained from COMSOL simulation and an excellent agreement is shown.
This method is expected to be very convenient for analytical investigation of various electromagnetic interactions of Terahertz (THz) waves (both free and guided) and charged particle bunches with slow-wave structures prospective in context of modern beam-driven THz emitters, THz accererators and THz-based bunch manipulation and bunch diagnostic systems.
\end{abstract}

\keywords{Diffraction radiation, open-ended waveguide}

\maketitle

\section{Introduction}

Cherenkov radiation (CR) has been initially discovered with fast electrons traversing dielectric meduim and emitting radiation in the visible region of electromagnetic spectrum~\cite{Ch37}. 
Through decades, CR has been succesfully used for a variety of applications in high-energy physics~\cite{Zrb}.
Today considerabrle advances have been reached in implementation of CR effect for dielectric wakefield acceleration~\cite{OShea16} where CR in the form of a wakefield with up to GeV per meter magnitude and Terahertz (THz) frequencies can be generated by high-quality relativistic electron bunches passing through dielectric-loaded waveguide structures (capillaries).

On the other hand, it has been far understood that radiation of the same nature occurs during the uniform movement of any localized source with the speed exceeding the phase velocity of electromagnetic waves in the given medium within the given frequency range~\cite{Askaryan1962}.
Again, Cherenkov-type radiation produced by short moving pulses of polarization generated in nonlinear crystals via optical rectification of laser pulses~\cite{Bass1962, Auston84} is considered nowadays as a most advanced and versatile mechanism for producing wideband THz radiation~\cite{BakunovBodrov2020} especially in the tilted-pulse-front scheme~\cite{Hebling2020, BodrovBakunov2019}.

In both mentioned areas we deal with powerful THz radiation which has (due to its unique properties) a huge amount of prospective applications, for example, those connected with possibilities of precise manipulating and probing the state of the matter~\cite{Hebling2020}.  
Moreover, as can be just noticed, contemporary beam and THz technologies has became tightly interlaced during last years. 
For example, strong THz fields allow realization of THz driven electron guns~\cite{Kartner2015}, performing THz bunch compression, streaking~\cite{AntipovXiang2020, Nanni2020} and wakefield acceleration within THz driven dielectric-lined waveguide structures~\cite{Nanni2015, Pacey2020}. 
Inversely, dielectric capillaries similar to those used for the THz bunch manipulation can be in turn utilized for development of high-power narrow-band THz sources~\cite{WangAntipov2018, GTAB14}.

It is worth noting that almost all the mentioned cases involve interaction of both THz waves (free or guided) and charged particle bunches with an open end of certain waveguide structure loaded with dielectric, most frequently a circular capillary~\cite{OShea16, AntipovXiang2020}.
For further development of the discussed prospective topics a rigorous approach allowing analytical investigation of both radiation from open-ended capillaries and their excitation by external source (bunch or electromagnetic pulse) would be very useful.
This is especially important, for example, for the case of high-order CR modes generation resulting in THz wakefields in mm-sized capillaries~\cite{GTAB14} since corresponding numerial simulations can be over-complicated.

However, up to now such an approach was absend despite the fact that general diffraction theory for open-ended wavegiude discontinuity was developed earlier%
~\cite{Weinb, Mittrab, KPV87, BGalst00, T14} but mainly for vacuum case.
The only known exception~\cite{VZh78} deals with dielectric-lined parallel-plate waveguides.
It is this old research which has inspired the investigation given in the present paper which fulfills the aforementioned gap in theory.

We present an elegant and efficient rigorous method for solving circular open-ended dielectric-loaded waveguide diffraction problems.
For the sake of clearness, we deal with the specific case of iniform dielectric loading and internal excitation by single waveguide mode, while layered loading and external excitation can be considered similarly.
Moreover, though the presented technique can be rigorously extended to the beam-driven case (similar to how it has been done for ``embedded'' structures~\cite{GTVGA19}), it can be applied approximately to the CR in the form of a narrow-band wakefield generated behind the driver bunch.
It is also notable that though the discussed approach is valid for the orthogonal waveguide end only whereas an inclined cut (Vlasov antenna) is often more convenient for practice, the obtained analytical results can be both extremely useful for improvement of corresponding approximate methods~\cite{GTAB14, IGTT14} and serve as reference points for numerical simulations.
 
\section{Problem formulation and general solution}

We consider a semi-infinite cylindrical waveguide with radius $a$ filled with a dielectric ($\varepsilon >1$) (Fig.~\ref{fig:geom}) and suppose that single $T{{M}_{0l}}$ waveguide mode incidents the orthogonal open end (cylindrical frame $\rho ,\varphi ,z$ is used):
%
\begin{figure}
\centering
\includegraphics[width=0.4\textwidth]{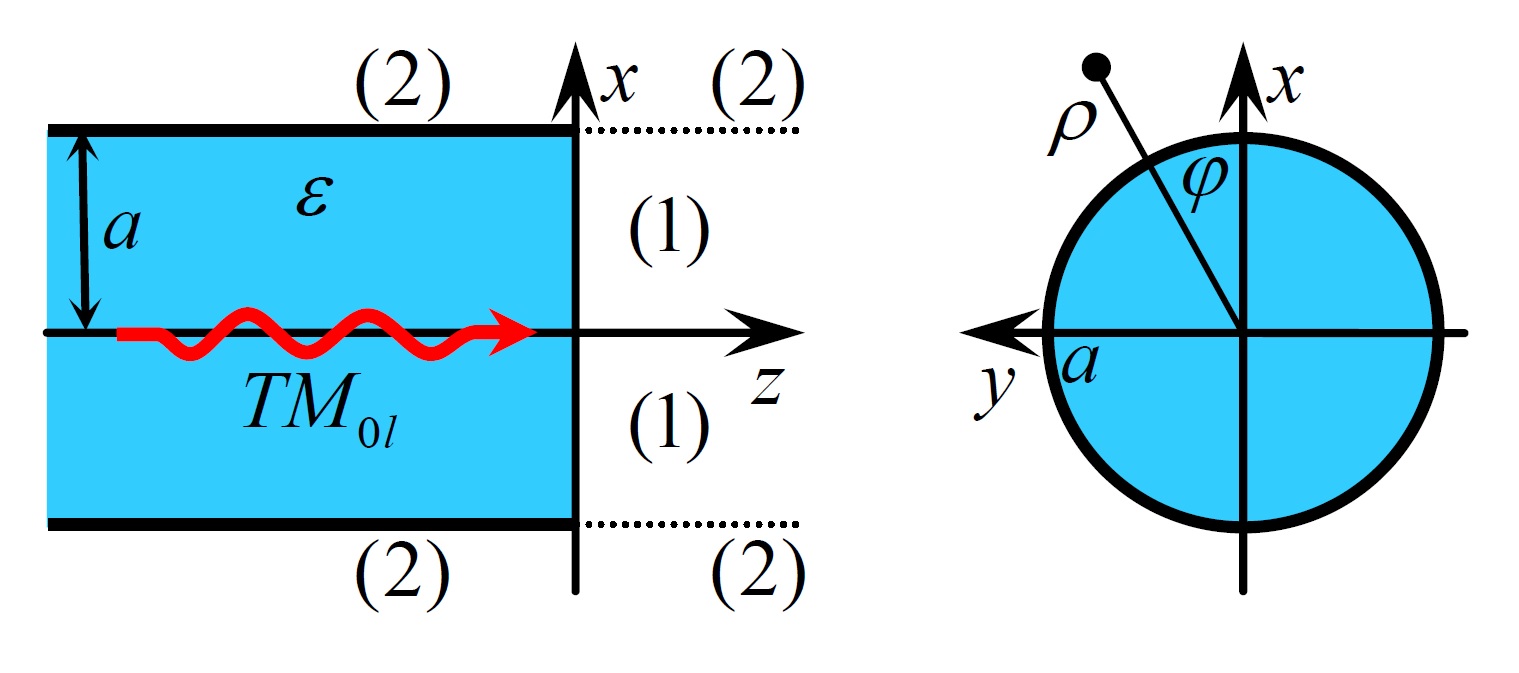}
\caption{\label{fig:geom} Geometry of the problem and main notations.}
\end{figure}
%
%
\begin{align}
H_{\omega \varphi }^{(i)}&=M^{(i)} {{J}_{1}}(\rho {{j}_{0l}}/a){{e}^{i{{k}_{zl}}z}}, \label{Hphii} \\
E_{ \omega \rho  } &= \frac{ 1 } {i k_0 \varepsilon } \frac{ \partial H_{\omega \varphi } }{\partial z },
\;
E_{ \omega z  } = \frac{ i } { k_0 \varepsilon }
\left( \frac{ H_{\omega \varphi } }{ \rho } + \frac{ \partial H_{\omega \varphi } }{\partial \rho } \right),
\end{align}
where
${{J}_{0}}({{j}_{0l}})=0$, ${{k}_{zl}}=\sqrt{k_{0}^{2}\varepsilon -j_{0l}^{2}{{a}^{-2}}}$,
$\operatorname{Im}{{k}_{zl}}>0$,
$ k_0 = \omega / c + i \delta$
(%
$ \delta \to 0 $,
which is equivalent to infinitely small dissipation an all areas).
Connection between~\eqref{Hphii} and CR wakefield generated by a charged particle bunch moving through the considered waveguide will be discussed below.
The reflected field in the area $z<0$, $\sqrt{{{x}^{2}}+{{y}^{2}}}=\rho <a$ is decomposed into a series of waveguide modes propagating in opposite direction:
\begin{equation}
H_{\omega \varphi }^{(r)}=\sum\nolimits_{m=1}^{\infty } M_m J_1 (\rho j_{0m}/a) e^{ -i k_{zm} z },
\end{equation}
where ${{M}_{m}}$ are unknown ``reflection coefficients'' that should be determined.
The vacuum area is divided into two subareas ``1'' and ``2'' (see Fig.~\ref{fig:geom}), where the field is described by Helmholtz equation:
\begin{equation}
\left[
{{{\partial }^{2}}}/{\partial {{z}^{2}}}+{{{\partial }^{2}}}/{\partial {{\rho }^{2}}}+{{\rho }^{-1}}{\partial }/{\partial \rho } +
\left( k_{0}^{2} - \rho^{-2} \right)
\right]
H_{\omega \varphi}^{(1,2)}=0.
\end{equation}
We introduce functions ${{\Psi }_{\pm }}(\rho ,\alpha )$ (hereafter subscripts $\pm $ mean that function is holomorphic and have no zeros in areas $\operatorname{Im}\alpha >-\delta$ and $\operatorname{Im}\alpha <\delta$, correspondingly):
\begin{equation}
\Psi _{+}^{(1,2)}(\rho ,\alpha )={{(2\pi )}^{-1}}\int_{0}^{\infty }{dzH_{\omega \varphi}^{(1,2)}(\rho ,z){{e}^{i\alpha z}}},
\end{equation}
%
\begin{equation}
\Psi _{-}^{(2)}(\rho ,\alpha )={{(2\pi )}^{-1}}\int_{-\infty }^{0}{dzH_{\omega \varphi}^{(2)}(\rho ,z){{e}^{i\alpha z}}},
\end{equation}
and similar transforms of
$ E_{\omega z}^{(1, 2)} $,
for example,
\begin{equation}
\Phi _{+}^{(1,2)}(\rho ,\alpha )={{(2\pi )}^{-1}}\int_{0 }^{\infty}{dz \frac{ k_0 }{ i } E_{\omega z}^{(1,2)}(\rho ,z){{e}^{i\alpha z}}}.
\end{equation}
From (4) we obtain
\begin{equation}\left( \frac{{{\partial }^{2}}}{\partial {{\rho }^{2}}} {+} \frac{1}{\rho }\frac{\partial }{\partial \rho } {+} {{\kappa }^{2}} {-} \frac{1}{\rho^2} \right)
\left\{ \begin{matrix}
   \Psi _{+}^{(1)}  \\
   \Psi _{-}^{(2)}{+}\Psi _{+}^{(2)}  \\
\end{matrix}
\right\}{=}
\left\{
\begin{matrix}
   {{F}^{(1)}}  \\
   0  \\
\end{matrix} \right\},
\end{equation}
%
\begin{equation}
2\pi{{F}^{(1)}} = {{\left. {\partial H_{\omega \varphi}^{(1)}}/{\partial z} \right|}_{z=+0}}-{{\left. i\alpha H_{\omega \varphi}^{(1)} \right|}_{z=+0}},
\end{equation}
where $\kappa =\sqrt{k_{0}^{2}-{{\alpha }^{2}}}$, $\operatorname{Im}\kappa >0$.
Function ${{F}^{(1)}}$ is determined using continuity of
$ E_{ \omega \rho } $ and $ H_{ \omega \varphi } $ at $z=0$, $\rho < a $, in the issue we obtain:
\begin{equation}
\label{gen_sol}
\begin{aligned}
&\Psi _{+}^{(1)}={{C}_{1}}{{J}_1}(\rho \kappa )+\Psi _{p}^{(1)}, \\
&\Psi _-^{(2)}+\Psi _+^{(2)}={{C}_{2}}H_{1}^{(1)}(\rho \kappa ),
\end{aligned}
\end{equation}
%
\begin{equation}
\Phi _{\pm}^{(1,2)} = \frac{ \Psi _{\pm}^{(1,2)} }{\rho}+\frac{ \partial \Psi _{\pm}^{(1,2)} }{\partial \rho},
\end{equation}
\begin{equation}
\begin{aligned}
\label{partsol}
& \Psi _{p}^{(1)}(\rho ,\alpha )=\frac{i}{2\pi }
\left[ M^{(i)}
\frac{ \frac{ k_{zl} }{\varepsilon} - \alpha }{ \alpha_l^2 -\alpha } J_{1} \left( \frac{\rho {{j}_{0l}}}{a} \right)-
\right. \\
&-\left. \sum\nolimits_{m=1}^{\infty }M_m
\frac{ \frac{ k_{zm} }{\varepsilon} + \alpha }{ \alpha_m^2 -\alpha } J_{1} \left( \frac{\rho {{j}_{0m}}}{a} \right)
 \right],
\end{aligned}
\end{equation}
where
$ \alpha_m = \sqrt{ k_0^2 - j_{ 0 m }^2 a^{-2} } $,
$ \mathrm{Im} \alpha_m >0 $
are longitudinal wavenumbers of vacuum waveguide,
${{C}_{1,2}}$ are unknown coefficients.
In particular, one obtains
$ \Phi _{+}^{(1)}(a, \alpha) = C_1 \kappa J_0( a \kappa) $,
$ \Phi _{+}^{(2)}(a, \alpha) = C_2 \kappa H_0^{(1)}( a \kappa) $
($\Phi _{-}^{(2)}(a, \alpha) = 0 $ because $ E_{ \omega z } = 0 $ for $ \rho =a $, $ z<0$), therefore
\begin{equation}
\label{C1C2}
\begin{aligned}
C_1&=\Phi _{+}^{(1)}( a, \alpha) \kappa^{-1} J_0^{-1}( a\kappa ), \\
C_2&=\Phi _{+}^{(2)}( a, \alpha) \kappa^{-1} \left( H_0^{(1)}( a\kappa ) \right)^{-1}.
\end{aligned}
\end{equation}
Note that
\begin{equation}
\label{cont}
\Psi _{+}^{(1)}(a, \alpha) = \Psi _{+}^{(2)}(a, \alpha),
\;
\Phi _{+}^{(1)}(a, \alpha) = \Phi _{+}^{(2)}(a, \alpha)
\end{equation}
due to continuity of
$ E_{ \omega z } $ and $ H_{ \omega \varphi } $ for
$ \rho = a $,
$ z > 0 $.
Using $C_1$ we get from~\eqref{gen_sol}, \eqref{partsol} and \eqref{cont}:
\begin{equation}
\begin{aligned}
&  \Psi _{+}^{(1)}(a ,\alpha )=
\frac{ J_1( a\kappa ) \Phi _{+}^{(2)}( a, \alpha) }{ \kappa J_0 ( a\kappa )} + \frac{i}{2\pi }  \times \\
& \times
\left[
M^{(i)} \frac{ \frac{ k_{zl} }{\varepsilon} - \alpha }{ \alpha_l^2 -\alpha } J_1 ( j_{ 0 l } ) -
\sum\limits_{m=1}^{\infty }M_m
\frac{ \frac{ k_{zm} }{\varepsilon} + \alpha }{ \alpha_m^2 -\alpha } J_1 ( j_{ 0 m } )
 \right].
\end{aligned}
\end{equation}
One can see that the function to the left of the ``=''~sign is regular in the area $ \mathrm{Im}\alpha>-\delta $ while the function to the right possesses pole singularity for $ \alpha = \alpha_p $, $ p = 1, 2, \ldots $ in this area.
Since the singularity at the right-hand side should be canceled, we obtain the following requirement:
\begin{equation}
\label{excludepoles}
\begin{aligned}
& \Phi _{+}^{(2)}(a,  \alpha_p )=\frac{ i a }{ 4 \pi } J_{1} ( j_{ 0 p } ) \times \\
& \times
\left[
\delta_{ l p } M^{(i)} \left( \frac{ k_{ z p } }{ \varepsilon } - \alpha_p \right) - M_p \left( \frac{ k_{ z p } }{ \varepsilon } + \alpha_p \right)
\right],
\end{aligned}
\end{equation}
where ${{\delta }_{l p}}$ is the Kronecker symbol.

Using general solution for the area ``2''~\eqref{gen_sol}, coefficient $C_2$~\eqref{C1C2} and continuity conditions~\eqref{cont} we arrive at the Wiener-Hopf-Fock equation:


%
\begin{equation}
\label{WHF1}
\begin{aligned}
&0 = \frac{ 2 i \Phi _{+}^{(2)}(a,\alpha ) }{ \kappa G(\alpha ) }+\Psi _-^{(2)}(a,\alpha ) + \frac{i}{2\pi }  \times \\
& {\times}
\left[
M^{(i)} \frac{ \frac{ k_{zl} }{\varepsilon} - \alpha }{ \alpha_l^2 -\alpha^2 } J_1 ( j_{ 0 l } ) -
\sum\limits_{m=1}^{\infty }M_m
\frac{ \frac{ k_{zm} }{\varepsilon} + \alpha }{ \alpha_m^2 -\alpha^2 } J_1 ( j_{ 0 m } )
 \right].
\end{aligned}
\end{equation}
where
$G(\alpha )=\pi a\kappa {{J}_{0}}(a\kappa )H_{0}^{(1)}(a\kappa )$.
Performing factorization,
$ \kappa = \kappa_+ \kappa_- $,
$ \kappa_{\pm} = \sqrt{ k_0 \pm \alpha } $,
$ G(\alpha ) = G_+(\alpha ) G_-(\alpha ) $, we obtain from~\eqref{WHF1} after multiplication by $\kappa_+ G_+$ and consequent decomposition of corresponding functions into a sum of ``+'' and ``--'' summands (standart formulas from~\cite{Mittrab} can be used):
\begin{equation}
\label{WHF2}
\begin{aligned}
&0 = \frac{ 2 \Phi _{+}^{(2)}(a,\alpha ) }{ \kappa_+G_+(\alpha ) }+\kappa_-G_-(\alpha )\Psi _-^{(2)}(a,\alpha ) +\times \\
& {\times} \frac{1}{2\pi }
\left[
M^{(i)} ( \eta_{l+}( \alpha )  + \eta_{l-}( \alpha )  ) J_1 ( j_{ 0 l } ) - \right. \\
& - \left.
\sum\limits_{m=1}^{\infty }M_m
( \zeta_{m+}( \alpha )  + \zeta_{m-}( \alpha )  ) J_1 ( j_{ 0 m } )
 \right],
\end{aligned}
\end{equation}
where
\begin{equation}
\label{etazeta}
\begin{aligned}
\eta_l(\alpha) &= \kappa_-(\alpha)G_-(\alpha) \frac{ \frac{ k_{zl} }{\varepsilon} - \alpha }{ \alpha_l^2 -\alpha^2 }, \\
\zeta_m(\alpha) &= \kappa_-(\alpha)G_-(\alpha) \frac{ \frac{ k_{zm} }{\varepsilon} + \alpha }{ \alpha_m^2 -\alpha^2 },
\end{aligned}
\end{equation}
\begin{equation}
\label{etazetaplus}
\begin{aligned}
\eta_{l+}(\alpha) &= \kappa_+(\alpha_l)G_+(\alpha_l) \frac{ \frac{ k_{zl} }{\varepsilon} + \alpha_l }{ 2\alpha_l (\alpha_l + \alpha ) }, \\
\zeta_{m+}(\alpha) &= \kappa_+(\alpha_m)G_+(\alpha_m) \frac{ \frac{ k_{zm} }{\varepsilon} - \alpha_m }{ 2 \alpha_m (\alpha_m +\alpha ) }.
\end{aligned}
\end{equation}
Equation~\eqref{WHF2} is solved in a common way: one should separate ``+'' and ``--'' terms into different parts of the equation:
\begin{equation}
\label{WHF3}
\begin{aligned}
&P(\alpha) = \frac{ 2 \Phi _{+}^{(2)}(a,\alpha ) }{ \kappa_+G_+(\alpha ) }+ \frac{ M^{(i)} \eta_{l+}( \alpha ) J_1 ( j_{ 0 l } ) }{2\pi }+ \\
& +\sum\limits_{m=1}^{\infty } \frac{ M_m \zeta_{m+}( \alpha ) J_1 ( j_{ 0 m } ) }{2\pi } = \frac{- M^{(i)} \eta_{l-}( \alpha ) J_1 ( j_{ 0 l } ) }{ 2 \pi } + \\
& + \sum\limits_{m=1}^{\infty } \frac{ M_m \zeta_{m-}( \alpha )  J_1 ( j_{ 0 m } ) }{ 2 \pi } - \kappa_- G_-(\alpha ) \Psi _-^{ ( 2 ) } ( a, \alpha ),
\end{aligned}
\end{equation}
where a polynomial function $ P(\alpha ) $ has been written based on analytic continuation theorem~%
\cite{Mittrab}.
To determine $P(\alpha)$ one should estimate asymptotic behaviour of all terms in~\eqref{WHF3} for $ |\alpha | \to \infty $, $-\delta<\mathrm{Im}\alpha < \delta$.
Based on Meixner edge condition~\cite{Mittrab} we have:
\begin{equation}
\label{Meixner}
\begin{aligned}
&\Phi _{+}^{(2)}(a,\alpha ) \underset{ |\alpha| \to \infty }{ \sim } \alpha^{-1/2-\tau}, \; \tau = \frac{1}{\pi} \mathrm{asin} \frac{\varepsilon-1}{2(\varepsilon+1)}, \\
& M_m\underset{ m \to \infty }{ \sim } m^{-1-\tau}, \; \Psi _-^{ ( 2 ) } ( a, \alpha ) \underset{ |\alpha| \to \infty }{ \sim } \alpha^{-3/2},
\end{aligned}
\end{equation}
therefore all terms in~\eqref{WHF3} decrase in accordance with power law and therefore $P(\alpha) = 0 $.
Formal solution of the Wiener-Hopf-Fock equation then reads
\begin{equation}
\label{WHFsol}
\begin{aligned}
&\Phi _{+}^{(2)}(a,\alpha ) = -\frac{ \kappa_+ ( \alpha ) G_+( \alpha ) }{ 4 \pi } \times \\
&\times
\left[
M^{(i)} \eta_{l+}( \alpha ) J_1 ( j_{ 0 l } ) +
\sum\limits_{m=1}^{\infty } M_m \zeta_{m+}( \alpha ) J_1 ( j_{ 0 m } )
\right].
\end{aligned}
\end{equation}
It should be noted that~\eqref{WHFsol} contains unknown coefficients $M_m$.
To resolve this, one should substitute~\eqref{WHFsol} into~\eqref{excludepoles}.
After that we obtain the following infinite linear system for $M_m$:
\begin{equation}
\label{sys}
\sum\nolimits_{m=1}^{\infty } W_{ p m } M_m = w_p,
\quad
p = 1, 2, \ldots,
\end{equation}
where
\begin{equation}
\label{Ww}
\begin{aligned}
& W_{ p m } =
J_1( j_{0 m} )
\left[
\zeta_{m+}( \alpha_p ) + \delta_{mp} i a \frac{ \frac{ k_{zm} }{\varepsilon} + \alpha_m }{ \kappa_+ ( \alpha_m ) G_+( \alpha_m ) }
\right], \\
& w_p =
M^{(i)} J_1( j_{0 l} )
\left[
\eta_{l+}( \alpha_p ) + \delta_{lp} i a \frac{ \frac{ k_{zl} }{\varepsilon} - \alpha_p }{ \kappa_+ ( \alpha_p ) G_+( \alpha_p ) }
\right].
\end{aligned}
\end{equation}
%
It can be easily shown that for $\varepsilon =1$ this system is analytically solved and the solution coincides with well-known result for open-ended vacuum waveguide~%
\cite{Weinb}.
For $\varepsilon \ne 1$ system~\eqref{sys} can be solved numerically using the reducing technique.

\section{Numerical results}

\begin{figure*}[t]
\centering
\includegraphics[width=0.95\textwidth]{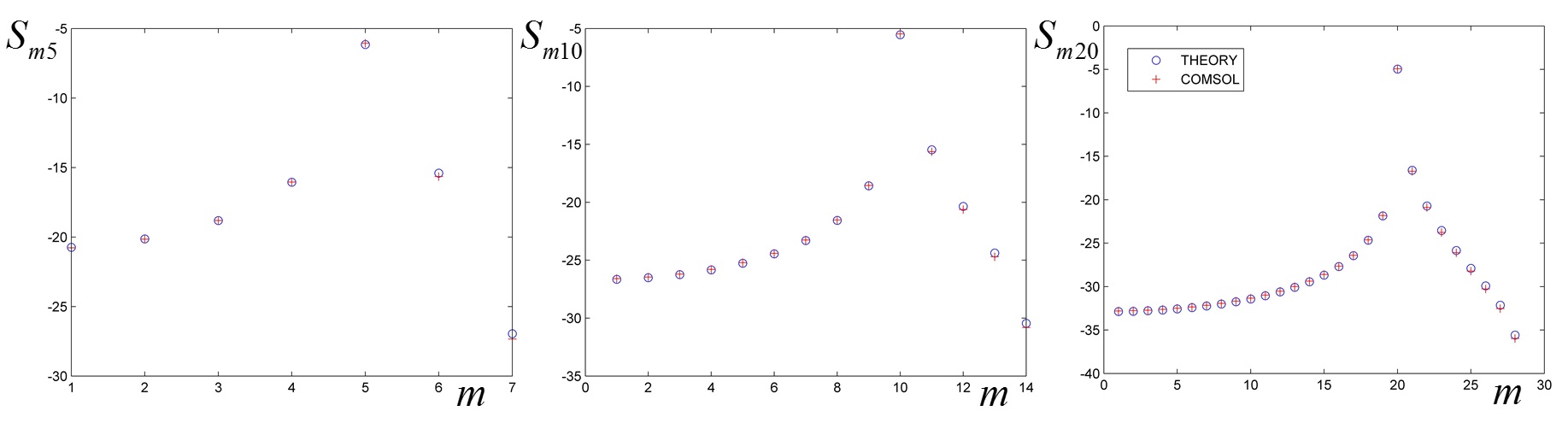}
	\caption{\label{fig:compare} Comparison between S-parameters (in dB) obtained via the presented analytical approach and via COMSOL simulations: $ S_{ml} $ corresponds to frequency  $ f^{\mathrm{CR}}_l $~\eqref{CHfreq} and incident mode with number $l$. We have 7 propagating modes for $ l = 5 $ ($f^{\mathrm{CR}}_5 = 300$~GHz), 14 for $ l = 10 $ ($f^{\mathrm{CR}}_{10} = 615$~GHz) and 28 for $ l = 20 $ ($f^{\mathrm{CR}}_{20} = 1.247$~THz). Other parameters: $ a = 0.24$~cm, $\varepsilon = 2$.}
\end{figure*}
%
Corresponding examples are the following.
For the case of $\varepsilon \ne 1$, we solve~\eqref{sys} by reducing it to the finite system of $ M_{\max } $ equations, where $ M_{\max } $ was chosen around 2-3 times as much as the total number of propagating modes in the waveguide at given frequency.
After that $ M_m $, $ m = 1,2, \ldots M_{\max} $ are immediately calculated, for example, in Matlab.
For convenient comparison between analytical results and results of numerical simulation, we have calculated powers carrying by incident mode and each reflected mode through the waveguide cross-section,
\begin{equation}
\label{powers}
\begin{aligned}
\Sigma^{ ( i ) } &= c a^2 / ( 8 k_0 \varepsilon ) J_{1}^{2} ( j_{ 0 l } )  \left| M^{(i)} \right|^{2} \mathrm{Re} ( k_{ z l } ), \\
\Sigma_{ m }^{ ( r ) } &= c a^{2} / ( 8 k_0 \varepsilon) J_{1}^{2} ( j_{0m} ) \left| M_{m} \right|^{2} \mathrm{Re} ( k_{ z m } ),
\end{aligned}
\end{equation}
and constructed corresponding $S$-parameters:
\begin{equation}
\label{spars}
S_{ m l } = \sqrt{ \left. \Sigma_{ m }^{ ( r ) } \right/ \Sigma^{ ( i ) } },
\end{equation}
which also can be expressed in dB, $S^{\mathrm{dB}}_{ m l} = 20 \mathrm{lg}S_{ m l }$.

Numerical simulations were performed in RF module of COMSOL Multiphysics package. The two dimensional frequency domain solver was utilized. An input end of the waveguide was supported by a series of numerical ports, one separate port for each propagating mode. The port which corresponds to the incident mode was set to be active and option ``active port feedback'' has been disabled. Corresponding eigenmodes were determined numerically, with analytically calculated longitudinal wavenumbers $k_{zm}$ being used as guess values. An open end of the waveguide was surrounded by a semisphere with scattering boundary condition applied. The length of the waveguide and the radius of damping semisphere radius were of the same order, at least several tens of maximum wavelength inside the waveguide.

For calculations presented below, the mode frequency was chosen to be equal to the frequency of CR mode $ f^{\mathrm{CR}}_l $ with numbers $l=5$, $l=10$ and $l=20$ produced by a moving charge having its Lorentz factor $\gamma = 7 $~%
\cite{GTAB14}:
\begin{equation}
\label{CHfreq}
\omega^{\mathrm{CR}}_l =
2 \pi f^{\mathrm{CR}}_l ={c\beta {{j}_{0l}}}/{\left( a\sqrt{\varepsilon {{\beta }^{2}}-1} \right)},
\end{equation}
where $\beta =\sqrt{1-{{\gamma }^{-2}}}$.
To clarify this choice of the frequency let us discuss the relation of the obtained results to the problem of diffraction of a charged paricle bunch wakefield at the open-end of the discussed dielectric-loaded waveguide.
Wakefield is a CR generated inside a wavegiude as an infinite set of discrete frequencies~\eqref{CHfreq}, while each frequency contribution to the total field is usually referred to as a CR mode.
A CR mode can be presented (after simplification) in the following form~\cite{B62}:
\begin{equation}
\label{CHmode}
H_{ \varphi l }^{\mathrm{CR}}
=
\mathrm{Im} \left[  i H_{\varphi 0 l } J_1 \left( \rho \frac{ j_{0l} }{a} \right) e^{ \frac{ i \omega^{\mathrm{CR}}_l z }{ c \beta } } e^{ - i \omega^{\mathrm{CR}}_l t } \right],
\end{equation}
where $ c \beta $ is bunch velocity, $H_{\varphi 0 l }$ is some constant.
Since for $ \omega = \omega^{\mathrm{CR}}_l $ we have $ k_{ z l } =  \omega^{\mathrm{CR}}_l  / ( c \beta ) $, an incident mode~\eqref{Hphii} corresponds to the $l$-th CR mode if $ M^{(i)} $ is chosen appropriately.

Figure~\ref{fig:compare} shows comparison between S-parameters calculated via presented rigorous analytical approach and obtained from COMSOL simulations.
As one can see, the agreement between results is excellent. This fact proves the presented theory and also shows correctness of COMSOL simulation procedure. One can see that typically the reflected mode with the number of incident mode dominates (it has the largest S-parameter), therefore the overall diffraction process is similar to a single mode reflection. However, other modes (especially those with close numbers) can be significant and therefore can alter mentioned ``close to single mode'' regime.

In conclusion, we have presented an elegant and convenient rigorous analytical approach for calculation of various diffraction processes at the open end (with orthogonal cut) of a circular waveguide with dielectric loading.
The obtained results have been compared to the results of simulations with commercial code COMSOL and an excellent agreement has been observed.
For simplicity and clearness of the presentation, in this short paper we have considered the problem with uniform dielectric filling.
However, a series of other more complicated problems which are closer to possible applications mentioned in the Introduction can be also considered using this powerful approach.
For example, excitation by a charged particle bunch (in full formulation including both wakefield and Coulomb field) or by an external electromagnetic wave can be incorporated into the solution and layered dielectric filling or corrugation of the waveguide wall can be investigated.
It is worth noting that computation resources used by the Matlab code based on analytical formulas is much smaller then those occupied by COMSOL.
For example, even for $ n = 20 $ and around $1$~THz frequency our code took about 30 seconds to calculate S-parameters shown in Fig.~\ref{fig:compare} (typical PC based on Intel Core i7 processor and Matlab Parallel Computing toolbox were used). COMSOL model took up to several hours to do the same task on a machine with similar processor, depending on used mesh.
Therefore, the discussed rigorous approach can be extremely useful for further development of various prospective applications based on electromagnetic interactions of single-cycle THz pulses, wide-band THz wakefields and charged particle bunches with dielectric-lined waveguide structures.

Besides mentioned open-ended waveguides with straight cut, this method can be also useful for investigation of structures with non-orthogonal cut.
Since in this case the rigorous theory can be marginally applied (solution for moreless similar parallel-plate dielectric-loaded waveguide problem has been reported just recently~\cite{Daniele2019}), development of reliable approximate methods is the most substantial idea (see, for example, \cite{GTAB14}).
Such reliable methods can be benchmarked and adjusted at simpler structures with orthogonal end cut.

\begin{acknowledgments}
This work was supported by Russian Science Foundation (Grant No. 18-72-10137).
S.N.G. is grateful to D. Minenkov, S. Simonov and S. Baturin for fruitful discussions.
\end{acknowledgments}


\begin{thebibliography}{26}%
\makeatletter
\providecommand \@ifxundefined [1]{%
 \@ifx{#1\undefined}
}%
\providecommand \@ifnum [1]{%
 \ifnum #1\expandafter \@firstoftwo
 \else \expandafter \@secondoftwo
 \fi
}%
\providecommand \@ifx [1]{%
 \ifx #1\expandafter \@firstoftwo
 \else \expandafter \@secondoftwo
 \fi
}%
\providecommand \natexlab [1]{#1}%
\providecommand \enquote  [1]{``#1''}%
\providecommand \bibnamefont  [1]{#1}%
\providecommand \bibfnamefont [1]{#1}%
\providecommand \citenamefont [1]{#1}%
\providecommand \href@noop [0]{\@secondoftwo}%
\providecommand \href [0]{\begingroup \@sanitize@url \@href}%
\providecommand \@href[1]{\@@startlink{#1}\@@href}%
\providecommand \@@href[1]{\endgroup#1\@@endlink}%
\providecommand \@sanitize@url [0]{\catcode `\\12\catcode `\$12\catcode
  `\&12\catcode `\#12\catcode `\^12\catcode `\_12\catcode `\%12\relax}%
\providecommand \@@startlink[1]{}%
\providecommand \@@endlink[0]{}%
\providecommand \url  [0]{\begingroup\@sanitize@url \@url }%
\providecommand \@url [1]{\endgroup\@href {#1}{\urlprefix }}%
\providecommand \urlprefix  [0]{URL }%
\providecommand \Eprint [0]{\href }%
\providecommand \doibase [0]{http://dx.doi.org/}%
\providecommand \selectlanguage [0]{\@gobble}%
\providecommand \bibinfo  [0]{\@secondoftwo}%
\providecommand \bibfield  [0]{\@secondoftwo}%
\providecommand \translation [1]{[#1]}%
\providecommand \BibitemOpen [0]{}%
\providecommand \bibitemStop [0]{}%
\providecommand \bibitemNoStop [0]{.\EOS\space}%
\providecommand \EOS [0]{\spacefactor3000\relax}%
\providecommand \BibitemShut  [1]{\csname bibitem#1\endcsname}%
\let\auto@bib@innerbib\@empty
\bibitem [{\citenamefont {\ifmmode~\check{C}\else
  \v{C}\fi{}erenkov}(1937)}]{Ch37}%
  \BibitemOpen
  \bibfield  {author} {\bibinfo {author} {\bibfnamefont {P.~A.}\ \bibnamefont
  {\ifmmode~\check{C}\else \v{C}\fi{}erenkov}},\ }\href {\doibase
  10.1103/PhysRev.52.378} {\bibfield  {journal} {\bibinfo  {journal} {Phys.
  Rev.}\ }\textbf {\bibinfo {volume} {52}},\ \bibinfo {pages} {378} (\bibinfo
  {year} {1937})}\BibitemShut {NoStop}%
\bibitem [{\citenamefont {Zrelov}(1970)}]{Zrb}%
  \BibitemOpen
  \bibfield  {author} {\bibinfo {author} {\bibfnamefont {V.~P.}\ \bibnamefont
  {Zrelov}},\ }\href@noop {} {\emph {\bibinfo {title} {Vavilov-Cherenkov
  Radiation in High-Energy Physics}}}\ (\bibinfo  {publisher} {Israel Program
  for Scientific Translations, Jerusalem},\ \bibinfo {year} {1970})\BibitemShut
  {NoStop}%
\bibitem [{\citenamefont {D.~O'Shea}\ \emph {et~al.}(2016)\citenamefont
  {D.~O'Shea}, \citenamefont {Andonian}, \citenamefont {Barber}, \citenamefont
  {Fitzmorris}, \citenamefont {Hakimi}, \citenamefont {Harrison}, \citenamefont
  {D.~Hoang}, \citenamefont {J.~Hogan}, \citenamefont {Naranjo}, \citenamefont
  {B.~Williams}, \citenamefont {Yakimenko},\ and\ \citenamefont
  {Rosenzweig}}]{OShea16}%
  \BibitemOpen
  \bibfield  {author} {\bibinfo {author} {\bibfnamefont {B.}~\bibnamefont
  {D.~O'Shea}}, \bibinfo {author} {\bibfnamefont {G.}~\bibnamefont {Andonian}},
  \bibinfo {author} {\bibfnamefont {S.}~\bibnamefont {Barber}}, \bibinfo
  {author} {\bibfnamefont {K.}~\bibnamefont {Fitzmorris}}, \bibinfo {author}
  {\bibfnamefont {S.}~\bibnamefont {Hakimi}}, \bibinfo {author} {\bibfnamefont
  {J.}~\bibnamefont {Harrison}}, \bibinfo {author} {\bibfnamefont
  {P.}~\bibnamefont {D.~Hoang}}, \bibinfo {author} {\bibfnamefont
  {M.}~\bibnamefont {J.~Hogan}}, \bibinfo {author} {\bibfnamefont
  {B.}~\bibnamefont {Naranjo}}, \bibinfo {author} {\bibfnamefont
  {O.}~\bibnamefont {B.~Williams}}, \bibinfo {author} {\bibfnamefont
  {V.}~\bibnamefont {Yakimenko}}, \ and\ \bibinfo {author} {\bibfnamefont
  {J.}~\bibnamefont {Rosenzweig}},\ }\href@noop {} {\bibfield  {journal}
  {\bibinfo  {journal} {Nature Communications}\ }\textbf {\bibinfo {volume}
  {7}},\ \bibinfo {pages} {12763} (\bibinfo {year} {2016})}\BibitemShut
  {NoStop}%
\bibitem [{\citenamefont {Askar'yan}(1962)}]{Askaryan1962}%
  \BibitemOpen
  \bibfield  {author} {\bibinfo {author} {\bibfnamefont {G.~A.}\ \bibnamefont
  {Askar'yan}},\ }\href@noop {} {\bibfield  {journal} {\bibinfo  {journal}
  {Soviet Physics -- JETP}\ }\textbf {\bibinfo {volume} {15}},\ \bibinfo
  {pages} {943} (\bibinfo {year} {1962})}\BibitemShut {NoStop}%
\bibitem [{\citenamefont {Bass}\ \emph {et~al.}(1962)\citenamefont {Bass},
  \citenamefont {Franken}, \citenamefont {Ward},\ and\ \citenamefont
  {Weinreich}}]{Bass1962}%
  \BibitemOpen
  \bibfield  {author} {\bibinfo {author} {\bibfnamefont {M.}~\bibnamefont
  {Bass}}, \bibinfo {author} {\bibfnamefont {P.~A.}\ \bibnamefont {Franken}},
  \bibinfo {author} {\bibfnamefont {J.~F.}\ \bibnamefont {Ward}}, \ and\
  \bibinfo {author} {\bibfnamefont {G.}~\bibnamefont {Weinreich}},\ }\href
  {\doibase 10.1103/PhysRevLett.9.446} {\bibfield  {journal} {\bibinfo
  {journal} {Phys. Rev. Lett.}\ }\textbf {\bibinfo {volume} {9}},\ \bibinfo
  {pages} {446} (\bibinfo {year} {1962})}\BibitemShut {NoStop}%
\bibitem [{\citenamefont {Auston}\ \emph {et~al.}(1984)\citenamefont {Auston},
  \citenamefont {Cheung}, \citenamefont {Valdmanis},\ and\ \citenamefont
  {Kleinman}}]{Auston84}%
  \BibitemOpen
  \bibfield  {author} {\bibinfo {author} {\bibfnamefont {D.~H.}\ \bibnamefont
  {Auston}}, \bibinfo {author} {\bibfnamefont {K.~P.}\ \bibnamefont {Cheung}},
  \bibinfo {author} {\bibfnamefont {J.~A.}\ \bibnamefont {Valdmanis}}, \ and\
  \bibinfo {author} {\bibfnamefont {D.~A.}\ \bibnamefont {Kleinman}},\ }\href
  {\doibase 10.1103/PhysRevLett.53.1555} {\bibfield  {journal} {\bibinfo
  {journal} {Phys. Rev. Lett.}\ }\textbf {\bibinfo {volume} {53}},\ \bibinfo
  {pages} {1555} (\bibinfo {year} {1984})}\BibitemShut {NoStop}%
\bibitem [{\citenamefont {Bakunov}\ \emph {et~al.}(2020)\citenamefont
  {Bakunov}, \citenamefont {Efimenko}, \citenamefont {Gorelov}, \citenamefont
  {Abramovsky},\ and\ \citenamefont {Bodrov}}]{BakunovBodrov2020}%
  \BibitemOpen
  \bibfield  {author} {\bibinfo {author} {\bibfnamefont {M.~I.}\ \bibnamefont
  {Bakunov}}, \bibinfo {author} {\bibfnamefont {E.~S.}\ \bibnamefont
  {Efimenko}}, \bibinfo {author} {\bibfnamefont {S.~D.}\ \bibnamefont
  {Gorelov}}, \bibinfo {author} {\bibfnamefont {N.~A.}\ \bibnamefont
  {Abramovsky}}, \ and\ \bibinfo {author} {\bibfnamefont {S.~B.}\ \bibnamefont
  {Bodrov}},\ }\href {\doibase 10.1364/OL.391871} {\bibfield  {journal}
  {\bibinfo  {journal} {Opt. Lett.}\ }\textbf {\bibinfo {volume} {45}},\
  \bibinfo {pages} {3533} (\bibinfo {year} {2020})}\BibitemShut {NoStop}%
\bibitem [{\citenamefont {Wang}\ \emph {et~al.}(2020)\citenamefont {Wang},
  \citenamefont {T\'oth}, \citenamefont {Hebling},\ and\ \citenamefont
  {K\"artner}}]{Hebling2020}%
  \BibitemOpen
  \bibfield  {author} {\bibinfo {author} {\bibfnamefont {L.}~\bibnamefont
  {Wang}}, \bibinfo {author} {\bibfnamefont {G.}~\bibnamefont {T\'oth}},
  \bibinfo {author} {\bibfnamefont {J.}~\bibnamefont {Hebling}}, \ and\
  \bibinfo {author} {\bibfnamefont {F.}~\bibnamefont {K\"artner}},\ }\href
  {\doibase 10.1002/lpor.202000021} {\bibfield  {journal} {\bibinfo  {journal}
  {Laser \& Photonics Reviews}\ }\textbf {\bibinfo {volume} {14}},\ \bibinfo
  {pages} {2000021} (\bibinfo {year} {2020})}\BibitemShut {NoStop}%
\bibitem [{\citenamefont {Bodrov}\ \emph {et~al.}(2019)\citenamefont {Bodrov},
  \citenamefont {Stepanov},\ and\ \citenamefont {Bakunov}}]{BodrovBakunov2019}%
  \BibitemOpen
  \bibfield  {author} {\bibinfo {author} {\bibfnamefont {S.~B.}\ \bibnamefont
  {Bodrov}}, \bibinfo {author} {\bibfnamefont {A.~N.}\ \bibnamefont
  {Stepanov}}, \ and\ \bibinfo {author} {\bibfnamefont {M.~I.}\ \bibnamefont
  {Bakunov}},\ }\href {\doibase 10.1364/OE.27.002396} {\bibfield  {journal}
  {\bibinfo  {journal} {Opt. Express}\ }\textbf {\bibinfo {volume} {27}},\
  \bibinfo {pages} {2396} (\bibinfo {year} {2019})}\BibitemShut {NoStop}%
\bibitem [{\citenamefont {Huang}\ \emph {et~al.}(2015)\citenamefont {Huang},
  \citenamefont {Nanni}, \citenamefont {Ravi}, \citenamefont {Hong},
  \citenamefont {Fallahi}, \citenamefont {Wong}, \citenamefont {Keathley},
  \citenamefont {Zapata},\ and\ \citenamefont {K\"artner}}]{Kartner2015}%
  \BibitemOpen
  \bibfield  {author} {\bibinfo {author} {\bibfnamefont {W.~R.}\ \bibnamefont
  {Huang}}, \bibinfo {author} {\bibfnamefont {E.~A.}\ \bibnamefont {Nanni}},
  \bibinfo {author} {\bibfnamefont {K.}~\bibnamefont {Ravi}}, \bibinfo {author}
  {\bibfnamefont {K.-H.}\ \bibnamefont {Hong}}, \bibinfo {author}
  {\bibfnamefont {A.}~\bibnamefont {Fallahi}}, \bibinfo {author} {\bibfnamefont
  {L.~J.}\ \bibnamefont {Wong}}, \bibinfo {author} {\bibfnamefont {P.~D.}\
  \bibnamefont {Keathley}}, \bibinfo {author} {\bibfnamefont {L.~E.}\
  \bibnamefont {Zapata}}, \ and\ \bibinfo {author} {\bibfnamefont {F.~X.}\
  \bibnamefont {K\"artner}},\ }\href {\doibase 10.1038/srep14899} {\bibfield
  {journal} {\bibinfo  {journal} {Scientific Reports}\ }\textbf {\bibinfo
  {volume} {5}},\ \bibinfo {pages} {14899} (\bibinfo {year}
  {2015})}\BibitemShut {NoStop}%
\bibitem [{\citenamefont {Zhao}\ \emph {et~al.}(2020)\citenamefont {Zhao},
  \citenamefont {Tang}, \citenamefont {Lu}, \citenamefont {Jiang},
  \citenamefont {Zhu}, \citenamefont {Hu}, \citenamefont {Song}, \citenamefont
  {Wang}, \citenamefont {Qiu}, \citenamefont {Jing}, \citenamefont {Antipov},
  \citenamefont {Xiang},\ and\ \citenamefont {Zhang}}]{AntipovXiang2020}%
  \BibitemOpen
  \bibfield  {author} {\bibinfo {author} {\bibfnamefont {L.}~\bibnamefont
  {Zhao}}, \bibinfo {author} {\bibfnamefont {H.}~\bibnamefont {Tang}}, \bibinfo
  {author} {\bibfnamefont {C.}~\bibnamefont {Lu}}, \bibinfo {author}
  {\bibfnamefont {T.}~\bibnamefont {Jiang}}, \bibinfo {author} {\bibfnamefont
  {P.}~\bibnamefont {Zhu}}, \bibinfo {author} {\bibfnamefont {L.}~\bibnamefont
  {Hu}}, \bibinfo {author} {\bibfnamefont {W.}~\bibnamefont {Song}}, \bibinfo
  {author} {\bibfnamefont {H.}~\bibnamefont {Wang}}, \bibinfo {author}
  {\bibfnamefont {J.}~\bibnamefont {Qiu}}, \bibinfo {author} {\bibfnamefont
  {C.}~\bibnamefont {Jing}}, \bibinfo {author} {\bibfnamefont {S.}~\bibnamefont
  {Antipov}}, \bibinfo {author} {\bibfnamefont {D.}~\bibnamefont {Xiang}}, \
  and\ \bibinfo {author} {\bibfnamefont {J.}~\bibnamefont {Zhang}},\ }\href
  {\doibase 10.1103/PhysRevLett.124.054802} {\bibfield  {journal} {\bibinfo
  {journal} {Phys. Rev. Lett.}\ }\textbf {\bibinfo {volume} {124}},\ \bibinfo
  {pages} {054802} (\bibinfo {year} {2020})}\BibitemShut {NoStop}%
\bibitem [{\citenamefont {Snively}\ \emph {et~al.}(2020)\citenamefont
  {Snively}, \citenamefont {Othman}, \citenamefont {Kozina}, \citenamefont
  {Ofori-Okai}, \citenamefont {Weathersby}, \citenamefont {Park}, \citenamefont
  {Shen}, \citenamefont {Wang}, \citenamefont {Hoffmann}, \citenamefont {Li},\
  and\ \citenamefont {Nanni}}]{Nanni2020}%
  \BibitemOpen
  \bibfield  {author} {\bibinfo {author} {\bibfnamefont {E.~C.}\ \bibnamefont
  {Snively}}, \bibinfo {author} {\bibfnamefont {M.~A.~K.}\ \bibnamefont
  {Othman}}, \bibinfo {author} {\bibfnamefont {M.}~\bibnamefont {Kozina}},
  \bibinfo {author} {\bibfnamefont {B.~K.}\ \bibnamefont {Ofori-Okai}},
  \bibinfo {author} {\bibfnamefont {S.~P.}\ \bibnamefont {Weathersby}},
  \bibinfo {author} {\bibfnamefont {S.}~\bibnamefont {Park}}, \bibinfo {author}
  {\bibfnamefont {X.}~\bibnamefont {Shen}}, \bibinfo {author} {\bibfnamefont
  {X.~J.}\ \bibnamefont {Wang}}, \bibinfo {author} {\bibfnamefont {M.~C.}\
  \bibnamefont {Hoffmann}}, \bibinfo {author} {\bibfnamefont {R.~K.}\
  \bibnamefont {Li}}, \ and\ \bibinfo {author} {\bibfnamefont {E.~A.}\
  \bibnamefont {Nanni}},\ }\href {\doibase 10.1103/PhysRevLett.124.054801}
  {\bibfield  {journal} {\bibinfo  {journal} {Phys. Rev. Lett.}\ }\textbf
  {\bibinfo {volume} {124}},\ \bibinfo {pages} {054801} (\bibinfo {year}
  {2020})}\BibitemShut {NoStop}%
\bibitem [{\citenamefont {Nanni}\ \emph {et~al.}(2015)\citenamefont {Nanni},
  \citenamefont {Huang}, \citenamefont {Hong}, \citenamefont {Ravi},
  \citenamefont {Fallahi}, \citenamefont {Moriena}, \citenamefont
  {Dwayne~Miller},\ and\ \citenamefont {K\"artner}}]{Nanni2015}%
  \BibitemOpen
  \bibfield  {author} {\bibinfo {author} {\bibfnamefont {E.~A.}\ \bibnamefont
  {Nanni}}, \bibinfo {author} {\bibfnamefont {W.~R.}\ \bibnamefont {Huang}},
  \bibinfo {author} {\bibfnamefont {K.-H.}\ \bibnamefont {Hong}}, \bibinfo
  {author} {\bibfnamefont {K.}~\bibnamefont {Ravi}}, \bibinfo {author}
  {\bibfnamefont {A.}~\bibnamefont {Fallahi}}, \bibinfo {author} {\bibfnamefont
  {G.}~\bibnamefont {Moriena}}, \bibinfo {author} {\bibfnamefont {R.~J.}\
  \bibnamefont {Dwayne~Miller}}, \ and\ \bibinfo {author} {\bibfnamefont
  {F.~X.}\ \bibnamefont {K\"artner}},\ }\href {\doibase 10.1038/ncomms9486}
  {\bibfield  {journal} {\bibinfo  {journal} {Nature Communications}\ }\textbf
  {\bibinfo {volume} {6}},\ \bibinfo {pages} {8486} (\bibinfo {year}
  {2015})}\BibitemShut {NoStop}%
\bibitem [{\citenamefont {Hibberd}\ \emph {et~al.}(2020)\citenamefont
  {Hibberd}, \citenamefont {Healy}, \citenamefont {Lake}, \citenamefont
  {Georgiadis}, \citenamefont {Smith}, \citenamefont {Finlay}, \citenamefont
  {Pacey}, \citenamefont {Jones}, \citenamefont {Saveliev}, \citenamefont
  {Walsh}, \citenamefont {Snedden}, \citenamefont {Appleby}, \citenamefont
  {Burt}, \citenamefont {Graham},\ and\ \citenamefont {Jamison}}]{Pacey2020}%
  \BibitemOpen
  \bibfield  {author} {\bibinfo {author} {\bibfnamefont {M.~T.}\ \bibnamefont
  {Hibberd}}, \bibinfo {author} {\bibfnamefont {A.~L.}\ \bibnamefont {Healy}},
  \bibinfo {author} {\bibfnamefont {D.~S.}\ \bibnamefont {Lake}}, \bibinfo
  {author} {\bibfnamefont {V.}~\bibnamefont {Georgiadis}}, \bibinfo {author}
  {\bibfnamefont {E.~J.~H.}\ \bibnamefont {Smith}}, \bibinfo {author}
  {\bibfnamefont {O.~J.}\ \bibnamefont {Finlay}}, \bibinfo {author}
  {\bibfnamefont {T.~H.}\ \bibnamefont {Pacey}}, \bibinfo {author}
  {\bibfnamefont {J.~K.}\ \bibnamefont {Jones}}, \bibinfo {author}
  {\bibfnamefont {Y.}~\bibnamefont {Saveliev}}, \bibinfo {author}
  {\bibfnamefont {D.~A.}\ \bibnamefont {Walsh}}, \bibinfo {author}
  {\bibfnamefont {E.~W.}\ \bibnamefont {Snedden}}, \bibinfo {author}
  {\bibfnamefont {R.~B.}\ \bibnamefont {Appleby}}, \bibinfo {author}
  {\bibfnamefont {G.}~\bibnamefont {Burt}}, \bibinfo {author} {\bibfnamefont
  {D.~M.}\ \bibnamefont {Graham}}, \ and\ \bibinfo {author} {\bibfnamefont
  {S.~P.}\ \bibnamefont {Jamison}},\ }\href {\doibase
  10.1038/s41566-020-0674-1} {\bibfield  {journal} {\bibinfo  {journal} {Nature
  Photonics}\ } (\bibinfo {year} {2020}),\
  10.1038/s41566-020-0674-1}\BibitemShut {NoStop}%
\bibitem [{\citenamefont {Wang}\ \emph {et~al.}(2018)\citenamefont {Wang},
  \citenamefont {Su}, \citenamefont {Du}, \citenamefont {Tian}, \citenamefont
  {Liang}, \citenamefont {Niu}, \citenamefont {Huang}, \citenamefont {Gai},
  \citenamefont {Yan}, \citenamefont {Tang},\ and\ \citenamefont
  {Antipov}}]{WangAntipov2018}%
  \BibitemOpen
  \bibfield  {author} {\bibinfo {author} {\bibfnamefont {D.}~\bibnamefont
  {Wang}}, \bibinfo {author} {\bibfnamefont {X.}~\bibnamefont {Su}}, \bibinfo
  {author} {\bibfnamefont {Y.}~\bibnamefont {Du}}, \bibinfo {author}
  {\bibfnamefont {Q.}~\bibnamefont {Tian}}, \bibinfo {author} {\bibfnamefont
  {Y.}~\bibnamefont {Liang}}, \bibinfo {author} {\bibfnamefont
  {L.}~\bibnamefont {Niu}}, \bibinfo {author} {\bibfnamefont {W.}~\bibnamefont
  {Huang}}, \bibinfo {author} {\bibfnamefont {W.}~\bibnamefont {Gai}}, \bibinfo
  {author} {\bibfnamefont {L.}~\bibnamefont {Yan}}, \bibinfo {author}
  {\bibfnamefont {C.}~\bibnamefont {Tang}}, \ and\ \bibinfo {author}
  {\bibfnamefont {S.}~\bibnamefont {Antipov}},\ }\href {\doibase
  10.1063/1.5042006} {\bibfield  {journal} {\bibinfo  {journal} {Review of
  Scientific Instruments}\ }\textbf {\bibinfo {volume} {89}},\ \bibinfo {pages}
  {093301} (\bibinfo {year} {2018})},\ \Eprint
  {http://arxiv.org/abs/https://doi.org/10.1063/1.5042006}
  {https://doi.org/10.1063/1.5042006} \BibitemShut {NoStop}%
\bibitem [{\citenamefont {Galyamin}\ \emph {et~al.}(2014)\citenamefont
  {Galyamin}, \citenamefont {Tyukhtin}, \citenamefont {Antipov},\ and\
  \citenamefont {Baturin}}]{GTAB14}%
  \BibitemOpen
  \bibfield  {author} {\bibinfo {author} {\bibfnamefont {S.~N.}\ \bibnamefont
  {Galyamin}}, \bibinfo {author} {\bibfnamefont {A.~V.}\ \bibnamefont
  {Tyukhtin}}, \bibinfo {author} {\bibfnamefont {S.}~\bibnamefont {Antipov}}, \
  and\ \bibinfo {author} {\bibfnamefont {S.~S.}\ \bibnamefont {Baturin}},\
  }\href {\doibase 10.1364/OE.22.008902} {\bibfield  {journal} {\bibinfo
  {journal} {Opt. Express}\ }\textbf {\bibinfo {volume} {22}},\ \bibinfo
  {pages} {8902} (\bibinfo {year} {2014})}\BibitemShut {NoStop}%
\bibitem [{\citenamefont {Weinstein}(1969)}]{Weinb}%
  \BibitemOpen
  \bibfield  {author} {\bibinfo {author} {\bibfnamefont {L.}~\bibnamefont
  {Weinstein}},\ }\href {https://books.google.ru/books?id=fbYOtAEACAAJ} {\emph
  {\bibinfo {title} {The Theory of Diffraction and the Factorization Method:
  generalized Wiener-Hopf Technique}}},\ Golem series in electromagnetics, v.
  3\ (\bibinfo  {publisher} {Golem Press},\ \bibinfo {year} {1969})\BibitemShut
  {NoStop}%
\bibitem [{\citenamefont {Mittra}\ and\ \citenamefont {Lee}(1971)}]{Mittrab}%
  \BibitemOpen
  \bibfield  {author} {\bibinfo {author} {\bibfnamefont {R.}~\bibnamefont
  {Mittra}}\ and\ \bibinfo {author} {\bibfnamefont {S.}~\bibnamefont {Lee}},\
  }\href@noop {} {\emph {\bibinfo {title} {Analytical Techniques in the Theory
  of Guided Waves}}}\ (\bibinfo  {publisher} {Macmillian},\ \bibinfo {year}
  {1971})\BibitemShut {NoStop}%
\bibitem [{\citenamefont {Kheifets}\ \emph {et~al.}(1987)\citenamefont
  {Kheifets}, \citenamefont {Palumbo},\ and\ \citenamefont {Vaccaro}}]{KPV87}%
  \BibitemOpen
  \bibfield  {author} {\bibinfo {author} {\bibfnamefont {S.}~\bibnamefont
  {Kheifets}}, \bibinfo {author} {\bibfnamefont {L.}~\bibnamefont {Palumbo}}, \
  and\ \bibinfo {author} {\bibfnamefont {V.~G.}\ \bibnamefont {Vaccaro}},\
  }\href {\doibase 10.1109/TNS.1987.4334809} {\bibfield  {journal} {\bibinfo
  {journal} {IEEE Transactions on Nuclear Science}\ }\textbf {\bibinfo {volume}
  {34}},\ \bibinfo {pages} {1094} (\bibinfo {year} {1987})}\BibitemShut
  {NoStop}%
\bibitem [{\citenamefont {Bolotovskii}\ and\ \citenamefont
  {Galst'yan}(2000)}]{BGalst00}%
  \BibitemOpen
  \bibfield  {author} {\bibinfo {author} {\bibfnamefont {B.~M.}\ \bibnamefont
  {Bolotovskii}}\ and\ \bibinfo {author} {\bibfnamefont {E.~A.}\ \bibnamefont
  {Galst'yan}},\ }\href {\doibase 10.3367/UFNr.0170.200008a.0809} {\bibfield
  {journal} {\bibinfo  {journal} {Usp. Fiz. Nauk}\ }\textbf {\bibinfo {volume}
  {170}},\ \bibinfo {pages} {809} (\bibinfo {year} {2000})}\BibitemShut
  {NoStop}%
\bibitem [{\citenamefont {Tyukhtin}(2014)}]{T14}%
  \BibitemOpen
  \bibfield  {author} {\bibinfo {author} {\bibfnamefont {A.~V.}\ \bibnamefont
  {Tyukhtin}},\ }\href {\doibase 10.1103/PhysRevSTAB.17.021303} {\bibfield
  {journal} {\bibinfo  {journal} {Phys. Rev. ST Accel. Beams}\ }\textbf
  {\bibinfo {volume} {17}},\ \bibinfo {pages} {021303} (\bibinfo {year}
  {2014})}\BibitemShut {NoStop}%
\bibitem [{\citenamefont {Voskresenskii}\ and\ \citenamefont
  {Zhurav}(1978)}]{VZh78}%
  \BibitemOpen
  \bibfield  {author} {\bibinfo {author} {\bibfnamefont {G.}~\bibnamefont
  {Voskresenskii}}\ and\ \bibinfo {author} {\bibfnamefont {S.}~\bibnamefont
  {Zhurav}},\ }\href@noop {} {\bibfield  {journal} {\bibinfo  {journal}
  {Radiotekhnika i Electronika}\ }\textbf {\bibinfo {volume} {23}},\ \bibinfo
  {pages} {2505} (\bibinfo {year} {1978})}\BibitemShut {NoStop}%
\bibitem [{\citenamefont {Galyamin}\ \emph {et~al.}(2019)\citenamefont
  {Galyamin}, \citenamefont {Tyukhtin}, \citenamefont {Vorobev}, \citenamefont
  {Grigoreva},\ and\ \citenamefont {Aryshev}}]{GTVGA19}%
  \BibitemOpen
  \bibfield  {author} {\bibinfo {author} {\bibfnamefont {S.~N.}\ \bibnamefont
  {Galyamin}}, \bibinfo {author} {\bibfnamefont {A.~V.}\ \bibnamefont
  {Tyukhtin}}, \bibinfo {author} {\bibfnamefont {V.~V.}\ \bibnamefont
  {Vorobev}}, \bibinfo {author} {\bibfnamefont {A.~A.}\ \bibnamefont
  {Grigoreva}}, \ and\ \bibinfo {author} {\bibfnamefont {A.~S.}\ \bibnamefont
  {Aryshev}},\ }\href {\doibase 10.1103/PhysRevAccelBeams.22.012801} {\bibfield
   {journal} {\bibinfo  {journal} {Phys. Rev. Accel. Beams}\ }\textbf {\bibinfo
  {volume} {22}},\ \bibinfo {pages} {012801} (\bibinfo {year}
  {2019})}\BibitemShut {NoStop}%
\bibitem [{\citenamefont {Ivanyan}\ \emph {et~al.}(2014)\citenamefont
  {Ivanyan}, \citenamefont {Grigoryan}, \citenamefont {Tsakanian},\ and\
  \citenamefont {Tsakanov}}]{IGTT14}%
  \BibitemOpen
  \bibfield  {author} {\bibinfo {author} {\bibfnamefont {M.}~\bibnamefont
  {Ivanyan}}, \bibinfo {author} {\bibfnamefont {A.}~\bibnamefont {Grigoryan}},
  \bibinfo {author} {\bibfnamefont {A.}~\bibnamefont {Tsakanian}}, \ and\
  \bibinfo {author} {\bibfnamefont {V.}~\bibnamefont {Tsakanov}},\ }\href
  {\doibase 10.1103/PhysRevSTAB.17.074701} {\bibfield  {journal} {\bibinfo
  {journal} {Phys. Rev. ST Accel. Beams}\ }\textbf {\bibinfo {volume} {17}},\
  \bibinfo {pages} {074701} (\bibinfo {year} {2014})}\BibitemShut {NoStop}%
\bibitem [{\citenamefont {Bolotovskii}(1962)}]{B62}%
  \BibitemOpen
  \bibfield  {author} {\bibinfo {author} {\bibfnamefont {B.~M.}\ \bibnamefont
  {Bolotovskii}},\ }\href {http://ufn.ru/ru/articles/1961/10/k/} {\bibfield
  {journal} {\bibinfo  {journal} {Physics-Uspekhi}\ }\textbf {\bibinfo {volume}
  {4}},\ \bibinfo {pages} {781} (\bibinfo {year} {1962})}\BibitemShut {NoStop}%
\bibitem [{\citenamefont {{Daniele}}\ \emph {et~al.}(2019)\citenamefont
  {{Daniele}}, \citenamefont {{Lombardi}},\ and\ \citenamefont
  {{Zich}}}]{Daniele2019}%
  \BibitemOpen
  \bibfield  {author} {\bibinfo {author} {\bibfnamefont {V.}~\bibnamefont
  {{Daniele}}}, \bibinfo {author} {\bibfnamefont {G.}~\bibnamefont
  {{Lombardi}}}, \ and\ \bibinfo {author} {\bibfnamefont {R.~S.}\ \bibnamefont
  {{Zich}}},\ }\href {\doibase 10.1109/TAP.2019.2948494} {\bibfield  {journal}
  {\bibinfo  {journal} {IEEE Transactions on Antennas and Propagation}\
  }\textbf {\bibinfo {volume} {67}},\ \bibinfo {pages} {7569} (\bibinfo {year}
  {2019})}\BibitemShut {NoStop}%
\end{thebibliography}
%

\end{document}